# Comparing CPU and GPU compute of PERMANOVA on MI300A

Igor Sfiligoi

University of California San Diego, San Diego Supercomputer Center, La Jolla, CA 92093, USA, isfiligoi@sdsc.edu

Comparing the tradeoffs of CPU and GPU compute for memory-heavy algorithms is often challenging, due to the drastically different memory subsystems on host CPUs and discrete GPUs. The AMD MI300A is an exception, since it sports both CPU and GPU cores in a single package, all backed by the same type of HBM memory. In this paper we analyze the performance of Permutational Multivariate Analysis of Variance (PERMANOVA), a non-parametric method that tests whether two or more groups of objects are significantly different based on a categorical factor. This method is memory-bound and has been recently optimized for CPU cache locality. Our tests show that GPU cores on the MI300A prefer the brute force approach instead, significantly outperforming the CPU-based implementation. The significant benefit of Simultaneous Multithreading (SMT) was also a pleasant surprise.

## 1 INTRODUCTION

The Permutational Multivariate Analysis of Variance (PERMANOVA) [1-2] is a popular method used in the microbiome field. It is is a non-parametric method that tests whether two or more groups of objects (e.g., samples) are significantly different based on a categorical factor, taking a distance matrix as its input. The statistical significance is assessed via a permutation test, with the assignment of objects to groups randomly permuted a number of times. This makes is extremely memory-intensive. The reference implementation is available in *scikit-bio* [3], with a further optimized version available in *unifrac-binaries* [4].

The AMD MI300A [5] is the first data-center Accelerated Processor Unit (APU) publicly available, and has been installed in several High-Performance Computing (HPC) centers. Unlike other compute setups, the MI300A combines both Central Processing Units (CPU) and Graphics Processing Units (GPU) in a single package, all of them paired with on-chip High Bandwidth Memory (HBM). This uniformity makes it a great target for comparing memory-heavy algorithmic performance between CPU and GPU implementations.

## 2 PERMANOVA INNER ALGORITHM

This section describes the most time-consuming part of the PERMANOVA algorithm, the parallel pseudo-F partial statistics computation (*permanova_f_stat_sW*). While the actual PERMANOVA includes several other steps that happen before and after, they add minimal overhead and will thus be ignored in this paper. The interested reader can find the complete source code at [6].

A typical PERMANOVA invocation uses a distance matrix between $1k^2$ and $100k^2$ elements, and computes the pseudo-F partial statistic on between 1k and 1M permutations. Each permutation is independent, making that dimension the most obvious parallelization target. The original, brute force implementation of PERMANOVA's parallel pseudo-F partial statistic is presented in Algorithm 1.

## ALGORITHM 1: Original brute force permanova_f_stat_sW

```
float permanova_f_stat_sW_one(mat[], n_dims, grouping[], inv_group_sizes[]) {
s_W = 0.0;
for (int row=0; row < (n_dims-1); row++) {  // no columns in last row
   for (int col=row+1; col < n_dims; col++) { // diagonal is always zero
     int group_idx = grouping[row];
     if (grouping[col] == group_idx) {
       const float * mat_row = mat + uint64_t(row)*uint64_t(n_dims);
       val = mat_row[col];
       s_W += val * val * inv_group_sizes[group_idx];;
     }
 }}
  return s_W;
}
void permanova_f_stat_sW_T(mat[], n_dims, groupings[], n_perms, inv_group_sizes[], group_sWs[]) {
#pragma omp parallel for
for (int p=0; p < n_perms; p++) {
  group_sWs[p] = permanova_f_stat_sW_one(mat, n_dims, groupings + p*n_dims, inv_group_sizes);
}
```

A careful look at the logic shows that the *grouping* array is accessed in a tiled manner, so the obvious step on CPU cores was to implement a tiled version of the algorithm. Unfortunately, many compilers do not support, or properly implement, the OpenMP *tile* directive on non-square nested loops, so we had to explicitly split the two loops by hand. As a side effect, we discovered that we could reuse the access to *inv_group_sizes* array in the innermost loop, saving both memory accesses and compute. The tiled version is available as Algorithm 2.

## ALGORITHM 2: Tiled permanova_f_stat_sW

```
float permanova_f_stat_sW_one(mat[], n_dims, grouping[], inv_group_sizes[]) {
s_W = 0.0;
for (int trow=0; trow < (n_dims-1); trow+=TILE) {  // no columns in last row
  for (int tcol=trow+1; tcol < n_dims; tcol+=TILE) { // diagonal is always zero
    for (uint32_t row=trow; row < min(trow+TILE,n_dims-1); row++) {
     min_col = std::max(tcol,row+1);  max_col = std::min(tcol+TILE,n_dims);
     float * mat_row = mat + row*n_dims;
     int group_idx = grouping[row];
     local_s_W = 0.0;
     for (uint32_t col=min_col; col < max_col; col++) {
       if (grouping[col] == group_idx) {
         val = mat_row[col];
         local_s_W += val * val;
       }
     }
     s_W += local_s_W*inv_group_sizes[group_idx];
 }}}
  return s_W;
}
```

```
void permanova_f_stat_sW_T(mat[], n_dims, groupings[], n_perms, inv_group_sizes[], group_sWs[]) {
#pragma omp parallel for
for (int p=0; p < n_perms; p++) {
  group_sWs[p] = permanova_f_stat_sW_one(mat, n_dims, groupings + p*n_dims, inv_group_sizes);
}
```

When it came time to port the code to GPU compute, we chose to use the GPU-offloading syntax of OpenMP [7], which required the insertion of only two pragma lines. Due to the much higher parallel nature of the GPUs, we parallelize both at permutation and distance matrix level there, as shown in Algorithm 3. Note that we ended up using the brute force version of the algorithm, as any attempt to tile the algorithm resulted in drastically slower execution.

ALGORITHM 3: GPU implementation of the brute force permanova_f_stat_sW

```
float permanova_f_stat_sW_one(mat[], n_dims, grouping[], inv_group_sizes[]) {
s_W = 0.0;
#pragma omp parallel for collapse(2) reduction(+:s_W)
for (int row=0; row < (n_dims-1); row++) {   // no columns in last row
  for (int col=row+1; col < n_dims; col++) { // diagonal is always zero
    int group_idx = grouping[row];
    if (grouping[col] == group_idx) {
      const float * mat_row = mat + uint64_t(row)*uint64_t(n_dims);
      val = mat_row[col];
      s_W += val * val * inv_group_sizes[group_idx];;
    }
 }}
  return s_W;
}
void permanova_f_stat_sW_T(mat[], n_dims, groupings[], n_perms, inv_group_sizes[], group_sWs[]) {
#pragma omp target teams distribute
for (int p=0; p < n_perms; p++) {
  group_sWs[p] = permanova_f_stat_sW_T_one(mat, n_dims, groupings + p*n_dims, inv_group_sizes);
}
```

## 3 BENCHMARK RESULTS ON MI300A

The above algorithms have been benchmarked as part of the early-user phase of the SDSC Cosmos system. Each Cosmos node contains 4 MI300A, but we used only one at a time, to contain the memory traffic within the single chip. The CPU code was compiled using gcc 14.2, while the GPU code was compiled using the clang-based AOMP 20.0 [8].

The input distance matrix had a size of $25145^2$ and was the result of computing the Unweighted Unifrac [9] on the Earth Microbiome Project (EMP) data. The number of permutations was 3999; the magnitude was chosen to be both large enough to exploit the GPU parallelism, small enough to result in reasonable execution time and would represent a realistic use case. The summary results for both CPU and GPU execution are available in Figure 1.

As can be seen, using the GPU compute units indeed results is drastically faster execution time. Compared to the same brute force algorithm on a non-SMT CPU setup, the GPU implementation is over 6x faster. That said, the more flexible nature of the CPU compute units allows for smarter algorithms, which claw back some of that advantage. This is especially noticeable when paired with Simultaneous Multithreading (SMT), that exposes two hardware threads for each physical CPU core.

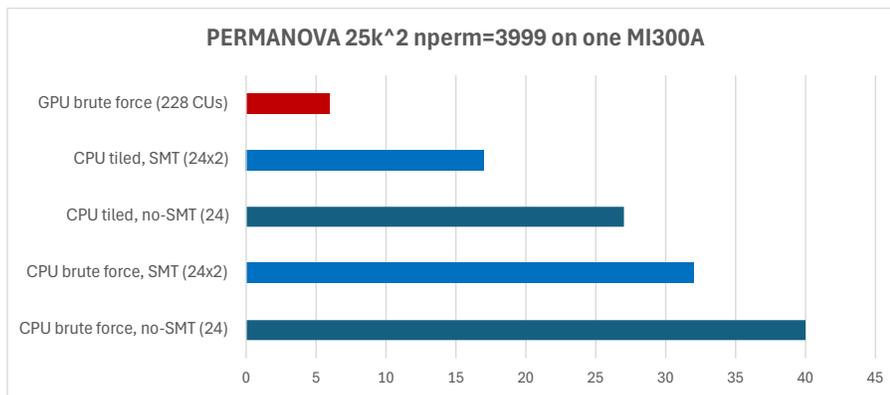

Figure 1: PERMANOVA execution time by algorithm and resource. The horizontal axis is expressed in seconds (lower is better).

In summary, even with an identical memory subsystem attached to both CPU and GPU compute units, massively parallel but memory-intensive algorithms like PERMANOVA do benefit from GPU compute. Some of the CPU-focused optimizations may however not directly translate to the GPU implementations, thus likely requiring some device-specific code.

**ACKNOWLEDGMENTS**

Data was collected as part of the early-user phase in the NSF-funded SDSC Cosmos system. All benchmark results should be considered preliminary, and results may change once the system is in production stage. This work was partially funded by the US National Research Foundation (NSF) under grants OAC-2404323 and DBI-2038509, and the US National Institutes of Health (NIH) grant R01CA241728.

## APPENDIX

### A1 Hardware details

Each Cosmos node contains 4 MI300A APUs. The tested nodes was in SPX mode, the user thus sees 192 logical cores and 4 logical GPUs, as outlined below:

```
> lscpu

Architecture:            x86_64
  CPU op-mode(s):        32-bit, 64-bit
  Address sizes:         48 bits physical, 48 bits virtual
  Byte Order:            Little Endian
CPU(s):                  192
  On-line CPU(s) list:   0-191
Vendor ID:               AuthenticAMD
  Model name:            AMD Instinct MI300A Accelerator
    CPU family:          25
    Model:               144
    Thread(s) per core:  2
    Core(s) per socket:  24
    Socket(s):           4
    Stepping:            1
    Frequency boost:     enabled
    CPU max MHz:         3700.0000
    CPU min MHz:         1500.0000
    BogoMIPS:            7400.37
…
Caches (sum of all):
  L1d:                   3 MiB (96 instances)
  L1i:                   3 MiB (96 instances)
  L2:                    96 MiB (96 instances)
  L3:                    384 MiB (12 instances)
NUMA:
  NUMA node(s):          4
  NUMA node0 CPU(s):     0-23,96-119
  NUMA node1 CPU(s):     24-47,120-143
  NUMA node2 CPU(s):     48-71,144-167
  NUMA node3 CPU(s):     72-95,168-191
…
> rocm-smi --showtoponuma

=========================== ROCm System Management Interface ===========================
====================================== Numa Nodes ======================================
GPU[0]          : (Topology) Numa Node: 0
GPU[0]          : (Topology) Numa Affinity: 0
GPU[1]          : (Topology) Numa Node: 1
GPU[1]          : (Topology) Numa Affinity: 1
GPU[2]          : (Topology) Numa Node: 2
GPU[2]          : (Topology) Numa Affinity: 2
GPU[3]          : (Topology) Numa Node: 3
GPU[3]          : (Topology) Numa Affinity: 3
================================== End of ROCm SMI Log ==================================
```

Only one APU was used during the tests, by setting

```
> export ROCR_VISIBLE_DEVICES=0
> taskset -c 0-23,96-119 <command>
```

**A2 Memory subsystem benchmarks**

The theoretical peak performance of each MI300A APU is 5.3 TB/s, as per AMD data sheet: https://www.amd.com/content/dam/amd/en/documents/instinct-tech-docs/data-sheets/amd-instinct-mi300a-data-sheet.pdf

The achievable memory throughput using a GPU-aware variant of the STREAM benchmark. The source code of modified STREAM benchmark test is available at https://github.com/sfiligoi/STREAM-OMPGPU . As can be seen, the GPU cores report approximately 3.0 TB/s, while the CPU cores report approximately 0.2 TB/s achievable memory throughput in the Triad test.

```
> export ROCR_VISIBLE_DEVICES=0
> taskset -c 0-23,96-119 ./stream.large.exe
-------------------------------------------------------------
STREAM version $Revision: 5.10 $
-------------------------------------------------------------
This system uses 8 bytes per array element.
-------------------------------------------------------------
Array size = 1000000000 (elements), Offset = 0 (elements)
Memory per array = 7629.4 MiB (= 7.5 GiB).
Total memory required = 22888.2 MiB (= 22.4 GiB).
Each kernel will be executed 10 times.
 The *best* time for each kernel (excluding the first iteration)
 will be used to compute the reported bandwidth.
-------------------------------------------------------------
Number of Threads requested = 48
Number of Threads counted = 48
-------------------------------------------------------------
Your clock granularity/precision appears to be 1 microseconds.
Each test below will take on the order of 51275 microseconds.
   (= 51275 clock ticks)
Increase the size of the arrays if this shows that
you are not getting at least 20 clock ticks per test.
-------------------------------------------------------------
WARNING -- The above is only a rough guideline.
For best results, please be sure you know the
precision of your system timer.
-------------------------------------------------------------
Function    Best Rate MB/s  Avg time     Min time     Max time
Copy:           199503.7     0.081749     0.080199     0.089379
Scale:          198570.4     0.080648     0.080576     0.080715
Add:            209086.6     0.116079     0.114785     0.120439
Triad:          209123.1     0.117878     0.114765     0.120415
-------------------------------------------------------------
Solution Validates: avg error less than 1.000000e-13 on all three arrays
-------------------------------------------------------------
```

```
> export ROCR_VISIBLE_DEVICES=0
> export HSA_XNACK=1
> taskset -c 0-23,96-119 ./stream.amd_apu.exe
-------------------------------------------------------------
STREAM version $Revision: 5.10 $
-------------------------------------------------------------
This system uses 8 bytes per array element.
-------------------------------------------------------------
Array size = 1000000000 (elements), Offset = 0 (elements)
Memory per array = 7629.4 MiB (= 7.5 GiB).
Total memory required = 22888.2 MiB (= 22.4 GiB).
Each kernel will be executed 10 times.
 The *best* time for each kernel (excluding the first iteration)
 will be used to compute the reported bandwidth.
-------------------------------------------------------------
Using accelerator
-------------------------------------------------------------
Your clock granularity/precision appears to be 1 microseconds.
Each test below will take on the order of 5797 microseconds.
   (= 5797 clock ticks)
Increase the size of the arrays if this shows that
you are not getting at least 20 clock ticks per test.
-------------------------------------------------------------
WARNING -- The above is only a rough guideline.
For best results, please be sure you know the
precision of your system timer.
-------------------------------------------------------------
Function    Best Rate MB/s  Avg time     Min time     Max time
Copy:           2981158.7     0.005496     0.005367     0.005596
Scale:          3056376.7     0.005385     0.005235     0.005466
Add:            3188574.5     0.007736     0.007527     0.007934
Triad:          3160344.6     0.007783     0.007594     0.007941
-------------------------------------------------------------
Solution Validates: avg error less than 1.000000e-13 on all three arrays
-------------------------------------------------------------
```